\RequirePackage[2020-02-02]{latexrelease}
\documentclass[eqsecnum,noshowpacs,showkeys,nofootinbib,aps]{revtex4}

\usepackage{amsmath}
\usepackage{epsfig}
\usepackage{xcolor}
\usepackage{ulem} 

\renewcommand{\theequation}{\arabic{equation}}
\def\be{\begin{equation}}
\def\ee{\end{equation}}
\def\bea{\begin{eqnarray}}
\def\eea{\end{eqnarray}}

\begin{document}

\title{Statistical Entropy Based on the Generalized-Uncertainty-Principle-Induced Effective Metric}
\author{Soon-Tae Hong}
\email{galaxy.mass@gmail.com} \affiliation{Center for Quantum
Spacetime and Department of Physics,
\\  Sogang University, Seoul 04107, Korea}
\author{Yong-Wan Kim}
\email{ywkim65@gmail.com}
 \affiliation{Department of Physics and
Research Institute of Physics and Chemistry, \\ Jeonbuk National
University, Jeonju 54896, Korea}
\author{Young-Jai Park}
\affiliation{Department of Physics, Sogang University, Seoul
04107, Korea }

\begin{abstract}
We investigate the statistical entropy of black holes within the
framework of the generalized uncertainty principle (GUP) by
employing effective metrics that incorporate leading-order and
all-orders quantum gravitational corrections. We construct three
distinct effective metrics induced by the GUP, which are derived
from GUP-corrected temperature, entropy, and all-orders GUP
corrections, and analyze their impact on black hole entropy using
't Hooft's brick wall method. Our results show that, despite the
differences in the effective metrics and the corresponding
ultraviolet cutoffs, the statistical entropy consistently
satisfies the Bekenstein-Hawking area law when expressed in terms
of an invariant (coordinate-independent) distance near the
horizon. Furthermore, we demonstrate that the GUP naturally
regularizes the ultraviolet divergence in the density of states,
eliminating the need for artificial cutoffs and yielding finite
entropy even when counting quantum states only in the vicinity of
the event horizon. These findings highlight the universality and
robustness of the area law under GUP modifications and provide new
insights into the interplay between quantum gravity effects and
black hole thermodynamics.
\end{abstract}





\maketitle

\section{Introduction}
\renewcommand{\theequation}{\arabic{section}.\arabic{equation}}

The generalized uncertainty principle (GUP) emerges from several
quantum gravity theories, including string theory
\cite{Amati:1987wq,Gross:1987kza,Amati:1988tn,Konishi:1989wk} and
loop quantum gravity
\cite{Ashtekar:1997yu,Smolin:2005re,Bojowald:2011jd,Bojowald:2020dkb}.
It suggests that the classic Heisenberg uncertainty principle
(HUP) needs to be modified at high energy scales or at extremely
small distances. A central tenet of many GUP formulations is the
existence of a minimal length
\cite{Maggiore:1993rv,Garay:1994en,Kempf:1994su,Kempf:1996nk,Scardigli:1999jh},
which is interpreted as a fundamental characteristic of quantum
gravity. This minimal length introduces inherent limits to
precision in position and momentum measurements
\cite{Adler:1999bu,Nicolini:2008aj,Das:2008kaa}.

In the context of black hole thermodynamics, the GUP plays a
crucial role in averting the complete evaporation of black holes,
thereby allowing for Planck-scale remnants that prevent
singularities
\cite{Adler:2001vs,Nozari:2005ah,Banerjee:2010sd,Dutta:2014yna}.
Without this intervention, small black holes would ultimately
evaporate completely due to Hawking radiation
\cite{Hawking:1975vcx}. Research utilizing frameworks like the
brick-wall model \cite{tHooft:1984kcu} has also suggested
potential modifications to black hole entropy. Furthermore,
employing a modified state density inspired by the GUP can
effectively eliminate divergences observed in the brick wall
model, eliminating the necessity for arbitrary cutoffs
\cite{Li:2002xb,Liu:2001ra,Liu:2003bb,Zhao:2003eu,Liu:2004xh,Kim:2006rx,
Nouicer:2007jg,Kim:2007if,Kim:2007nh,Eune:2010kx,Anacleto:2014apa,
Tang:2017wph,Vagenas:2019wzd,Li:2021ohz,Hong:2021xeg,Li:2023mix}.

Various forms of GUP have been proposed, including those with
linear and quadratic terms in momentum or even higher-order terms,
some of which also predict a maximal observable momentum
\cite{Kempf:1994su,Nouicer:2007jg,Nouicer:2007pu,Pedram:2011gw,Tawfik:2014zca,
Chung:2019raj,Petruzziello:2020een,Du:2022mvr,Hemeda:2022dnd,Sonnino:2024cxe}.
To analyze the effects of GUP in curved spacetime, particularly
around black holes, researchers often employ the concept of an
effective metric
\cite{Scardigli:2014qka,Contreras:2016xib,Anacleto:2020lel,
Ong:2023jkp,Hong:2024xog,Ong:2025ent}, which encodes quantum
gravitational corrections into a modified spacetime geometry. This
approach is highly effective because the thermodynamic properties
of black holes, such as entropy and temperature, are intrinsically
tied to the spacetime metric (e.g., via surface gravity and
horizon area). By constructing GUP-modified effective metrics, one
can systematically study how quantum gravity alters black hole
thermodynamics while retaining the mathematical framework of
classical general relativity. For instance, the brick wall method
for entropy calculation relies on the near-horizon metric
structure, and GUP-induced corrections to the metric naturally
regularize ultraviolet divergences without ad hoc cutoffs. This
synergy between effective metrics and thermodynamic principles
underscores how spacetime geometry serves as a bridge between
quantum gravity phenomenology and observable black hole
properties.

Despite its utility as a tool, the construction and application of
GUP-based effective metrics face significant challenges. There is
no universally accepted method for deriving the effective metric
from a given GUP. Different assumptions or approaches can lead to
different metric forms, raising questions about their validity and
uniqueness. Concerns have also been raised that using series
truncations in derivations rather than full GUP-corrected
expressions can lead to incorrect results or artificial
singularities \cite{Ong:2023jkp}. This challenges the necessity of
using full GUP expressions in quantum gravity phenomenology.

{In this paper, we present a systematic and unified analysis of
black hole statistical entropy in the framework of the GUP by
constructing and comparing three distinct GUP-induced effective
metrics: those derived from the leading-order GUP-corrected
temperature, the leading-order GUP-corrected entropy, and the
all-order GUP-corrected temperature. One of the main findings in
this paper is that, despite differences in the effective metrics
and corresponding ultraviolet cutoffs, all cases yield the same
invariant distance near the horizon and lead to a universal
recovery of the Bekenstein--Hawking area law for black hole
entropy. Furthermore, we demonstrate that the GUP itself provides
a natural regularization of the ultraviolet divergence in the
density of states, eliminating the need for arbitrary cutoffs.
This clarifies the universality of the area law under quantum
gravity corrections and highlights the robustness and physical
significance of the effective metric approach in understanding
black hole thermodynamics.}

The remainder of the paper is organized as follows: in
Section~\ref{s2}, we introduce the GUP, addressing both the
leading order and all orders in Planck length. In
Section~\ref{s3}, we derive effective metrics based on the
leading-order GUP-corrected temperature and entropy, as well as
from the GUP-corrected temperature considering all orders in
Planck length. In Section~\ref{s4}, we compute the free energy and
entropy using the brick wall method based on the effective metrics
identified in the previous section. This analysis demonstrates
that the area laws are satisfied and that all three effective
metrics remarkably yield the same invariant distance. In
Section~\ref{s5}, we again calculate the free energy and entropy
by carefully counting the number of quantum states near the event
horizon without any artificial cutoff. Then, we compare the
results with those from Section~\ref{s4}. Finally, in
Section~\ref{s6}, we provide our conclusions.

\section{GUP to Leading and All Orders in the Planck Length}
\label{s2}

The most common form of the GUP is expressed as
 \be\label{gup}
 \Delta x \Delta p \geq \frac{\hbar}{2}\left(1+\alpha L_{p}^{2}
                        \frac{\Delta p^{2}}{\hbar^2}\right).
 \ee
Here, $\Delta x$ and $\Delta p$ represent the uncertainties in
position and momentum, respectively. The reduced Planck constant
is denoted by $\hbar$, and the Planck length is denoted by $L_p$,
which represents an extremely short distance at which quantum
gravitational effects become significant. $\alpha$ is a
dimensionless GUP parameter, typically of order unity. If the GUP
parameter $\alpha$ vanishes (i.e., $\alpha=0$), this equation
reverts to the standard HUP.

Solving the above GUP inequality for the momentum uncertainty
$\Delta p$ yields the following range:
 \be\label{gup-leading-order}
 \frac{\hbar}{\alpha L^2_p}\Delta x\left(1-\sqrt{1-\frac{\alpha L^2_p}{\Delta x^{2}}}\right) \leq \Delta p
 \leq \frac{\hbar}{\alpha L^2_p}\Delta x\left(1+\sqrt{1-\frac{\alpha L^2_p}{\Delta x^{2}}}\right).
 \ee
A crucial consequence of this equation is that the term inside the
square root must be positive. This condition imposes a minimum
value on the position uncertainty $\Delta x$. This minimum
measurable length is given by
 \be\label{minL}
 (\Delta x)_{\rm min}=\sqrt{\alpha}L_p.
 \ee
This implies that GUP suggests the existence of a fundamental
minimum length in space, a significant departure from standard
quantum mechanics.

With growing interest in quantum phenomenology involving GUP with
higher orders in the Planck length, Nouicer \cite{Nouicer:2007jg}
generalized the GUP in Equation (\ref{gup}) to include all orders
in Planck length. This all-order GUP correction in the Planck
length is expressed as
 \be\label{all-order-gup}
 \Delta x \Delta p \geq \frac{\hbar}{2}e^{\alpha L^2_p\frac{\Delta p^2}{\hbar^2}}.
 \ee
This expression can be reduced to the original GUP in Equation
(\ref{gup}) if we only consider the leading order in the Planck
length. {By squaring the GUP in Equation (\ref{all-order-gup}), we
arrive at the following {inequality}:
 \be
 -2\alpha L^2_p\frac{\Delta p^2}{\hbar^2}e^{-2\alpha L^2_p\frac{\Delta p^2}{\hbar^2}}
 {\le} -\frac{\alpha L^2_p}{2\Delta x^2}.
 \ee
Defining $W(\xi)$ and $\xi$ as
 \be
 W(\xi)\equiv-2\alpha L^2_p\frac{\Delta p^2}{\hbar^2}, ~~~ \xi\equiv-\frac{\alpha L^2_p}{2\Delta x^2},
 \ee
it can be shown that the GUP for all orders in the Planck length
satisfies
 \be\label{Lambert}
 W(\xi)e^{W(\xi)}{\le}~\xi,
 \ee
{where $W(\xi)$ is a multi-valued Lambert function
\cite{Corless:1996zz}.} For the range $-1/e\le\xi\le 0$, it has
two real values, $W_0(\xi)$ and $W_{-1}(\xi)$. For $\xi\ge 0$, it
has one real value, $W_0(\xi)$. Here, $W_0(\xi)$ denotes the
principal branch satisfying $W(\xi) \ge -1$, and $W_{-1}(\xi)$
denotes the branch satisfying $W(\xi)\le -1$. The branch point
{occurs} at $\xi=-1/e$ {and} provides a minimum length for the GUP
to all orders in the Planck length:
 \be\label{minLAO}
 \Delta x \ge \sqrt{\frac{e}{2}}\cdot\sqrt{\alpha}L_p \equiv (\Delta x)_{\rm min}.
 \ee
The uncertainty in momentum from Equation  (\ref{all-order-gup})
can be expressed using the Lambert $W$ function as
 \be\label{mom-all-order}
 \Delta p {\ge} \frac{\hbar}{2\Delta x}e^{-\frac{1}{2}W(-\frac{\alpha L^2_p}{2\Delta x^2})}.
 \ee
Throughout this paper, unless otherwise stated, we adopt natural
units in which the Planck length ($L_p$) and the reduced Planck
constant ($\hbar$) are set to unity ($L_P=\hbar=1$).

\section{GUP-Induced Effective Metric}\label{s3}

Following a proposal by Ong \cite{Ong:2023jkp}, we will briefly
review the methods for finding effective metrics from the leading
order GUP in the Planck length and then extend these methods to
the all-order GUP in the Planck length. An effective metric
represents a modified metric tensor that describes how GUP
influences the geometry of spacetime.

\subsection{Effective Metric from the Leading-Order GUP-Corrected Temperature}\label{s31}

According to Adler, Chen, and Santiago (ACS) \cite{Adler:2001vs},
by assuming that photons escape from a Schwarzschild black hole at
its event horizon (radius $r_H=2M$, where $M$ is the black hole's
mass) and that the spectrum of these escaping photons is thermal,
one can derive the Hawking temperature for an asymptotically flat
Schwarzschild black hole. The generalized momentum $\Delta p$ is
identified with the characteristic energy of the Hawking
particles, $E=k_BT=pc$ (where $k_B$ is the Boltzmann constant, $T$
is temperature, and $c$ is the speed of light), and the
generalized position $\Delta x$ is identified with the horizon
size $r_H$. Then, from the lower bound of the generalized momentum
uncertainty in Equation  (\ref{gup-leading-order}), the Hawking
temperature incorporating the GUP effect can be obtained as
 \be\label{gupT}
 T_{\mathrm{GUP}}=\frac{M}{\pi \alpha}
 \left(1-\sqrt{1-\frac{\alpha}{4M^{2}}}\right).
 \ee
A factor of $1/2\pi$ has been introduced here so that as
$\alpha\rightarrow 0$, we recover the standard Hawking temperature
of a Schwarzschild black hole, $T_{\rm Sch}=1/8\pi M$.

To incorporate GUP into an effective metric, one can consider a
metric ansatz, without loss of generality, of the form
 \be\label{metric}
 d s^{2}=-f(r) d t^{2}+f(r)^{-1} d r^{2}+r^{2} d \Omega^{2}.
 \ee
Here, $f(r)$ is a function of the radial coordinate $r$, and
$d\Omega^2$ is the line element of a unit 2-sphere. Assuming the
areal radius remains intact (see discussion in
\cite{Ong:2023jkp}), $f(r)$ can be written as
 \be\label{mansatz}
 {f(r)=f_S(r) g(r),~~{\rm with}~~ f_S(r)\equiv\left(1-\frac{2M}{r}\right).}
 \ee
This form of ansatz has the advantage that it modifies the Hawking
temperature simply by a proportionality to the function $g(r)$
 \be\label{gupT-ansatz}
 T=\frac{1}{4\pi}\left.f'(r)\right|_{r=r_H}=\frac{g(r_H)}{8\pi M}.
 \ee
Equating this temperature with the GUP-induced temperature
(\ref{gupT}), one obtains
 \be
 g(r_H)=\frac{2r^2_H}{\alpha}\left(1-\sqrt{1-\frac{\alpha}{r^2_H}}\right).
 \ee
Thus, the explicit form of $g(r)$ can be inferred as
 \be
 g(r)=\frac{2r^2}{\alpha}\left(1-\sqrt{1-\frac{\alpha}{r^2}}\right).
 \ee
This leads to the following candidate for the effective metric
function $f_A(r)$ corresponding to the leading-order GUP corrected
temperature as
 \be\label{eff-metric1}
 f_A(r)=\left(1-\frac{2M}{r}\right)\frac{2r^2}{\alpha}\left(1-\sqrt{1-\frac{\alpha}{r^2}}\right).
 \ee

\subsection{Effective Metric from the Leading-Order GUP-Corrected Entropy}

The entropy from the GUP temperature (\ref{gupT}) can be found by
integrating $1/T_{GUP}$ with respect to $M$
 \bea
 S_{\rm GUP}&=&\int\frac{dM}{T_{\rm GUP}}\nonumber\\
  &=&2\pi M^2\left(1+\sqrt{1-\frac{\alpha}{4M^2}}\right)
    -\frac{\pi\alpha}{2}\ln\left(M+M\sqrt{1-\frac{\alpha}{4M^2}}\right),
 \eea
up to some constant terms that depend only on $\alpha$. For large
$M$, this entropy approximates to
 \be\label{ent-series}
 S_{\rm GUP}\simeq 4\pi M^2
 -\frac{\pi\alpha}{4}\ln(4\pi M^2)+\frac{\pi^2\alpha^2}{16}(4\pi M^2)^{-1}
 +\frac{\pi^3\alpha^3}{64}(4\pi M^2)^{-2}+\cdots.
 \ee

Another way to obtain an effective metric is to start with the
series-expanded entropy (\ref{ent-series})}, noting that $dS_{\rm
GUP}/dM=1/T$, where $T$ is given by Equation (\ref{gupT-ansatz}).
Then, in the lowest order in $\alpha$, one has
 \be
 8\pi M\left(1-\frac{\alpha}{16M^2}\right)=\frac{8\pi M}{g(r_H)}.
 \ee
Thus, one can infer that
 \be
 g(r)=\left(1-\frac{\alpha}{4r^2}\right)^{-1}.
 \ee
And we finally arrive at the effective metric function $f_B(r)$
 \be\label{eff-metric2}
 f_B(r)=\left(1-\frac{2 M}{r}\right)\left(1-\frac{\alpha}{4r^2}\right)^{-1}.
 \ee
Note that compared with Equation  (\ref{eff-metric1}), this
effective metric has a curvature singularity at
$r=\sqrt{\alpha}/2$ \cite{Contreras:2016xib} but one in Equation
(\ref{eff-metric1}) at $r=\sqrt{\alpha}$
\cite{Ong:2023jkp,Hong:2024xog}.

\subsection{Effective Metric from the All-Order GUP-Corrected Temperature}

As in the Section III.A, according to ACS, the spectrum of
escaping photons that satisfy the all-order GUP correction gives
us the corresponding GUP temperature as
 \be
 T_{\rm GUP}=\frac{1}{4\pi r_H}e^{-\frac{1}{2}W(-\frac{\alpha}{2r^2_H})}.
 \ee
Comparing this with the temperature in Equation
(\ref{gupT-ansatz}) from the metric ansatz (\ref{mansatz}), one
can infer
 \be
 g(r)=e^{-\frac{1}{2}W(-\frac{\alpha}{2r^2})}.
 \ee
As a result, one can arrive at an effective metric function
$f_C(r)$ as
 \be\label{eff-metric3}
 f_C(r)=\left(1-\frac{2M}{r}\right)e^{-\frac{1}{2}W(-\frac{\alpha}{2r^2})}.
 \ee
This is the effective metric corrected by the GUP to all orders in
the Planck length.

Finally, for later use, we summarize the surface gravities
$\kappa_H(=\frac{1}{2}\frac{df}{dr}\mid_{r=r_H})$ corresponding to
the effective metrics as follows
 \begin{eqnarray}
 \kappa_S&=&{\left.\frac{1}{2}\frac{df_S}{dr}\right|_{r=r_H}}=\frac{1}{4M},\\
 \kappa_A&=&{\left.\frac{1}{2}\frac{df_A}{dr}\right|_{r=r_H}}=\frac{2M}{\alpha}\left(1-\sqrt{1-\frac{\alpha}{4M^2}}\right),\\
 \kappa_B&=&{\left.\frac{1}{2}\frac{df_B}{dr}\right|_{r=r_H}}=\frac{1}{4M\left(1-\frac{\alpha}{16M^2}\right)},\\
 \kappa_C&=&{\left.\frac{1}{2}\frac{df_C}{dr}\right|_{r=r_H}}=\frac{e^{-\frac{1}{2}W(-\frac{\alpha}{2r^2})}}{4M}\label{kappaC}.
 \end{eqnarray}
Here, $f_S$, $f_A$, $f_B$, and $f_C$ are the original
Schwarzschild metric (\ref{mansatz}) and the effective metrics
(\ref{eff-metric1}), (\ref{eff-metric2}) and (\ref{eff-metric3})
for the leading-order GUP-corrected temperature, the leading-order
GUP-corrected entropy and the all-order GUP-corrected temperature,
respectively.
Note also that when the GUP parameter $\alpha$ is very small, they
become approximately
 \begin{eqnarray}
 \kappa_A&\simeq&\frac{1}{4M}+\frac{\alpha}{64M^3}+\frac{\alpha^2}{512M^5}+{\cal O}(\alpha^3),\\
 \kappa_B&\simeq&\frac{1}{4M}+\frac{\alpha}{64M^3}+\frac{\alpha^2}{1024M^5}+{\cal O}(\alpha^3),\\
 \kappa_C&\simeq&\frac{1}{4M}+\frac{\alpha}{64M^3}+\frac{5\alpha^2}{2048M^5}+{\cal O}(\alpha^3),
 \end{eqnarray}
respectively, except for the pure Schwarzschild metric. As a
result, one can see that
 \be
 \kappa_S<\kappa_B<\kappa_A<\kappa_C.
 \ee

\section{Revisiting the Brick Wall Model Using Effective Metrics}\label{s4}

The statistical entropy of black holes, as derived using 't
Hooft's brick wall method, arises from the analysis of quantum
fields in curved spacetime. Starting with the Klein--Gordon
equation for a scalar field $\Phi$ as
 \be\label{KGeq}
 (\nabla^2-m^2)\Phi=0,
 \ee
the radial modes are decomposed as $\Phi=e^{-i\omega
t}\varphi(r)Y(\theta,\phi)$, where $Y(\theta,\phi)$ are spherical
harmonics. Quantizing these modes yields the free energy $F$ for
bosonic fields
 \be\label{freeE}
 F=-\frac{1}{\pi}\int^{\infty}_0\frac{g(\omega)}{e^{\beta\omega}-1} d\omega,
 \ee
where $g(\omega)$ represents the density of states derived from
the metric by
 \be\label{nos}
 g(\omega)=\frac{2}{3}\int^{r_1}_{r_0} \frac{r^2}{f^{1/2}(r)}\left(\frac{\omega^2}{f(r)}-m^2\right)^{3/2}dr.
 \ee
The statistical entropy is then given by
 \be
 S=\beta^2\left.\frac{\partial F}{\partial\beta}\right|_{\beta=\beta_H},
 \ee
where $\beta_H=2\pi/\kappa_H$.


The original brick wall method imposes boundary conditions
 \be
 \varphi(r_H+h)=\varphi(L)=0.
 \ee
Here, $\varphi(r)$ is the radial wave function, $h$ is a UV cutoff
near the horizon, and $L$ is an IR cutoff confining the system.
These cutoffs are necessary to regulate divergences in the density
of states.

The free energy simplifies under a small mass approximation to the
term
 \be
 F\simeq -\frac{2\pi^3}{45\beta^4}\int^L_{r_H+h}\frac{r^2}{f^2(r)}dr,
 \ee
to be integrated. Then, for the Schwarzschild black hole, this
leads to
 \be
 F_S \simeq  -\frac{2\pi^3}{45\beta^4}\left(\frac{16M^4}{h_S}+\frac{L^3}{3}+32M^3\log\frac{L}{h_S}\right),
 \ee
where the last two terms are the contribution from the vacuum
surrounding the system at large distance. The $1/h_S$ term
dominates because $h_S\ll L$, making vacuum contributions
negligible. Therefore, we have
 \be
 F_S\simeq -\frac{32\pi^3M^4}{45\beta^4h_S},
 \ee
and can find the entropy as
 \be\label{ori-brickwall-entropy}
 S_S \simeq \frac{1}{720\pi Mh_S}\left(\frac{A}{4}\right).
 \ee
Here, $A=16\pi M^2$ is the area of the event horizon. By choosing
the cutoff $h_S$ as
 \be\label{hcutoff}
 h_S=\frac{1}{720\pi M},
 \ee
the statistical entropy matches the Bekenstein--Hawking entropy
 \be
 S=\frac{A}{4}.
 \ee

It is important to note that the brick wall cutoff $h$ is
coordinate-dependent, which is considered a coordinate artifact.
To address this, we introduce the invariant distance given by
 \be
 l_{\rm inv}=\int^{r_H+h_S}_{r_H}\frac{dr}{\sqrt{f(r)}}\simeq 2\sqrt{2Mh_S}.
 \ee
Then, the entropy can be rewritten in terms of this invariant
distance, allowing for a mass-independent, constant cutoff while
preserving the area law
 \be
 S=\frac{1}{90\pi l^2_{\rm inv}}\left(\frac{A}{4}\right).
 \ee
Therefore, if we choose the invariant distance as
 \be\label{inv_dis0}
 l_{\rm inv}=\frac{1}{\sqrt{90\pi}},
 \ee
we have the mass-independent constant cutoff while keeping the
area law intact.

As we have seen in the previous sections, modifications from the
GUP alter the effective metric, introducing $\alpha$-dependent
terms. For the effective metric derived from the leading-order
GUP-corrected temperature, one can find the free energy as
 \begin{eqnarray}
 F_A \simeq &-&\frac{2\pi^3}{45\beta^4}
         \left(\frac{8M^4(1+\sqrt{1-\frac{\alpha}{4M^2}})-M^2\alpha}{h_A}
         +\frac{L^3}{6}\left(1-\sqrt{1-\frac{\alpha}{L^2}}\right)\right.\nonumber\\
         &+& \left.16M\left(M^2-\frac{\alpha}{16}
         +\frac{2M^3-\frac{3}{8}M\alpha}{\sqrt{4M^2-\alpha}}\right)\log\frac{L}{h_A}\right).
 \end{eqnarray}
Then, from the first term,  after ignoring the vacuum contributing
the last two terms, the entropy can be obtained as
 \be
 S_A\simeq \frac{1}{360\pi M h_A(1+\sqrt{1-\frac{\alpha}{4M^2}})}
           \left(\frac{A}{4}\right).
 \ee
Therefore, if we choose the cutoff $h_A$ as
 \be\label{ha}
 h_A = \frac{1}{360\pi M(1+\sqrt{1-\frac{\alpha}{4M^2}})},
 \ee
the statistical entropy recovers the area law in black hole
thermodynamics. Moreover, the coordinate artifact, the brick wall
cutoff $h_A$ can be replaced by the invariant distance
 \be
 l_{A,\rm inv}=\frac{1}{\sqrt{90\pi}},
 \ee
by following the same procedure as before. This is the same
invariant distance as the one of the original Schwarzschild black
hole. Note that the entropy $S_A$ and the cutoff $h_A$ recover the
correct limits of Equations (\ref{ori-brickwall-entropy}) and
(\ref{hcutoff}) as $\alpha\rightarrow 0$, respectively. And the
invariant distance $l^2_{A,\rm inv}$ remains the same with
Equation (\ref{inv_dis0}).

For the effective metric from the leading-order GUP-corrected
entropy, the free energy is
 \be
 F_B\simeq -\frac{2\pi^3}{45\beta^4}
         \left(\frac{16(M^2-\frac{\alpha}{16})^2}{h_B}
         +\frac{L^3}{3}+32M\left(M^2-\frac{\alpha}{16}\right)\log\frac{L}{h_B}\right).
 \ee
Ignoring the vacuum contributing the last two terms, as before,
the entropy from the first term is given by
 \be
 S_B\simeq \frac{1}{720\pi M h_B(1-\frac{\alpha}{16M^2})}
           \left(\frac{A}{4}\right).
 \ee
Therefore, if we choose the cutoff $h_B$ as
 \be\label{hb}
 h_B = \frac{1}{720\pi M(1-\frac{\alpha}{16M^2})},
 \ee
the statistical entropy recovers the area law in black hole
thermodynamics. Moreover, the brick wall cutoff $h_B$ can be
replaced by the invariant distance
 \be
 l_{B,\rm inv}=\frac{1}{\sqrt{90\pi}},
 \ee
by following the same procedure as before. Note that the entropy
$S_B$ and the cutoff $h_B$ recover the correct limits of Equations
(\ref{ori-brickwall-entropy}) and (\ref{hcutoff}), respectively,
as $\alpha\rightarrow 0$.

Finally, for the effective metric from the all-order GUP-corrected
temperature, the free energy is approximately
 \be
 F_C\simeq -\frac{2\pi^3}{45\beta^4}
         \left(16M^3 e^{W(-\frac{\alpha}{8M^2})}\left(\frac{M}{h_C}+\log\frac{L}{h_C}\right)\right).
 \ee
After ignoring the vacuum contributing the last term, as before,
the entropy from the first term is given by
 \be
 S_C\simeq \frac{e^{-\frac{1}{2}W(-\frac{\alpha}{8M^2})}}{720\pi M h_C}
           \left(\frac{A}{4}\right).
 \ee
Therefore, if we choose the cutoff $h_C$ as
 \be\label{hc}
 h_C = \frac{e^{-\frac{1}{2}W(-\frac{\alpha}{8M^2})}}{720\pi M},
 \ee
the statistical entropy recovers the area law in the black hole
thermodynamics. Moreover, the brick wall cutoff $h_C$ can be
replaced by the invariant distance
 \be
 l_{C,\rm inv}=\frac{1}{\sqrt{90\pi}},
 \ee
by following the same procedure as before. Note that the entropy
$S_C$ and the cutoff $h_C$ recover the correct limits of Equations
(\ref{ori-brickwall-entropy}) and (\ref{hcutoff}), respectively,
as $\alpha\rightarrow 0$.

Table \ref{evth} summarizes the cutoffs and invariant distances
for the GUP-corrected effective metrics of the Schwarzschild black
hole. The UV cutoffs near the event horizon are apparently
different from each other; however, interestingly, the invariant
distances are the same in all   cases. One can compare their
relative differences by expanding in $\alpha$ as
 \begin{eqnarray}
 h_A &\simeq& \frac{1}{720\pi M}+\frac{\alpha}{11520\pi M^3}+\frac{\alpha^2}{92160\pi M^5}+{\cal O}(\alpha^3),\nonumber\\
 h_B &\simeq& \frac{1}{720\pi M}+\frac{\alpha}{11520\pi M^3}+\frac{\alpha^2}{184320\pi M^5}+{\cal O}(\alpha^3),\nonumber\\
 h_C &\simeq& \frac{1}{720\pi M}+\frac{\alpha}{11520\pi M^3}+\frac{\alpha^2}{73728\pi M^5}+{\cal O}(\alpha^3),
 \end{eqnarray}
showing that
 \be
 h_S < h_B < h_A < h_C.
 \ee

\begingroup
\setlength{\tabcolsep}{6pt} 
\renewcommand{\arraystretch}{1.8}
\begin{table}[ht!]
  \begin{center}
    \caption{Effective metrics, cutoff $h$ formula, and invariant distances $l_{\rm inv}$}
    \label{evth}
    \begin{tabular}{|c|c|c|c|}
    \hline
       metric  & cutoff $h$ formula  & invariant distance $l_{\rm inv}$ \\
    \hline
   Schwarzschild & $h_S=\frac{1}{720\pi M}$      &    $\frac{1}{\sqrt{90\pi}}$   \\
      $f_A(r)$ & $h_A = \frac{1}{360\pi M(1+\sqrt{1-\frac{\alpha}{4M^2}})}$  &   $\frac{1}{\sqrt{90\pi}}$        \\
      $f_B(r)$ & $h_B = \frac{1}{720\pi M(1-\frac{\alpha}{16M^2})}$      &      $\frac{1}{\sqrt{90\pi}}$  \\
      $f_C(r)$ & $h_C = \frac{e^{-\frac{1}{2}W(-\frac{\alpha}{8M^2})}}{720\pi M}$      &      $\frac{1}{\sqrt{90\pi}}$   \\
    \hline
    \end{tabular}
  \end{center}
\end{table}
\endgroup

\section{Statistical Entropy Based on the GUP Induced Effective Metrics}\label{s5}

Let us calculate the statistical entropy of a free scalar field on
the Schwarzschild black hole with the effective metrics
considering the near-horizon contributions of quantum states.
First of all, it is well-known that when the gravity is turned on,
the number of quantum states in a volume element in phase space
are changed from $(2\pi)^3$ into $(2\pi)^3(1+\alpha p^2)^3$ for
the leading-order GUP correction, and $(2\pi)^3 e^{\alpha p^2}$
for the all-order GUP correction, respectively
\cite{Li:2002xb,Zhao:2003eu,Liu:2004xh,Kim:2006rx}. Specifically,
in (3 + 1) dimensions, they are
\begin{equation}
\label{dnL} dn = \frac{d^3 x d^3 p}{(2\pi)^3(1+\alpha p^2)^3},
\end{equation}
for the leading-order GUP corrections, and
\begin{equation}
\label{dnA} dn = \frac{d^3 x d^3 p}{(2\pi)^3}e^{-\alpha p^2},
\end{equation}
for the all-order GUP correction. Note that both the leading-order
GUP-corrected temperature and entropy approaches yield the same
phase space modification, as given in Equation (\ref{dnL}). The
square module of momentum $p^2$ is given by
\begin{equation}
\label{smom}  p^{2}\equiv g^{rr}{p_{r}}^{2} + g^{\theta
\theta}{p_{\theta}}^{2}+ g^{\phi\phi}{p_{\phi}}^{2} =
\frac{\omega^{2}}{f} - \mu^{2}.
\end{equation}
When $\alpha\rightarrow 0$, they are reduced to the usual number
of quantum states in HUP \cite{tHooft:1984kcu}.

Now, substituting the ansatz of the wave function
$\Phi(t,r,\theta, \phi) = e^{-i\omega t}\psi(r, {\theta}, \phi)$
in the Klein--Gordon Equation (\ref{KGeq}), we have
\begin{equation}
\label{rtheta0}
 \partial_{r}^2 \psi + \left( \frac{f'}{f} + \frac{2}{r}\right)\partial_{r} \psi
 + \frac{1}{f}\left({\frac{1}{r^2}}\left[\partial^2_\theta + {\rm cot}\theta
\partial_\theta + {\frac{1}{{\rm sin}^{2}\theta}}\partial^2_\phi \right] +
\frac{\omega^{2}}{f} - \mu^{2} \right)\psi = 0,
\end{equation}
where $f'$ denotes the differentiation with respect to $r$. By
using the Wenzel--Kramers--Brillouin approximation
\cite{tHooft:1984kcu} with $\psi \sim e^{iS(r,\theta,\phi)}$ and
keeping the real parts, we have the following modified dispersion
relation
\begin{equation}
\label{wkb}
 p_\mu p^\mu = - \frac{\omega^2}{f} + f{p_{r}}^{2} + \frac{p^2_\theta}{r^2}
   + \frac{p_{\phi}^2}{r^2 {\rm sin}^2 \theta}=-\mu^2,
\end{equation}
where $p_{r} = \frac{\partial S}{\partial r}$, $p_{\theta} =
\frac{\partial S}{\partial \theta}$ and  $p_{\phi} =
\frac{\partial S}{\partial \phi}$. Then, one can easily calculate
the volume of the momentum phase space as
\begin{eqnarray}
V_{p}(r,\theta)= \int dp_{r}dp_{\theta}dp_{\phi} =\frac{4\pi}{3}
\frac{r^2 {\rm sin}\theta}{\sqrt{f}}
\left(\frac{\omega^2}{f}-\mu^2 \right)^{\frac{3}{2}},
\end{eqnarray}
which satisfy $\omega\geq\mu\sqrt{f}$.

\subsection{Leading-Order GUP Correction}

From Equations (\ref{dnL}) and (\ref{smom}), the number of quantum
states related to the radial mode with energy less than $\omega$
is given by
\begin{eqnarray}
 \label{TnqsL}
 n(\omega) = \int dn
 = \frac{2}{3\pi}\int_{r_H} dr
   \frac{r^2\left(\frac{{\omega}^2}{f}-\mu^{2}\right)^{\frac{3}{2}}}{\sqrt{f}\left(1+\alpha(\frac{\omega^2}{f}-\mu^2)\right)^3}.
\end{eqnarray}
For the bosonic case, the free energy of a thermal ensemble of
scalar fields at inverse temperature $\beta$ is given by
\begin{eqnarray}
\label{TfreeEL}
 F&=& \frac{1}{\beta}\sum_K \ln \left( 1 - e^{-\beta \omega_K}
 \right)\nonumber\\
     &=& - \frac{2}{3\pi} \int_{r_H} dr \frac{r^2}{\sqrt{f}}
   \int^{\infty}_{\mu\sqrt{f}} d\omega~
    \frac{\left(\frac{{\omega}^2}{f}- \mu^{2}\right)^{\frac{3}{2}}}{(e^{\beta \omega} -1)\left(1+\alpha(\frac{\omega^2}{f}-\mu^2)\right)^3}.
\end{eqnarray}
Here, we have considered the continuum limit, integrated it by
parts, and used the number of quantum states in Equation
(\ref{TnqsL}).

Now, we are only interested in the contribution from just the
vicinity near the event horizon in the range of $(r_H, r_H +
\epsilon)$ where $\epsilon$ is the brick wall cutoff used to
remove ultraviolet divergences. Since $f \rightarrow 0$ near the
event horizon,  the $\frac{{\omega}^2}{f}$ term is dominant in
$\frac{{\omega}^2}{f}-\mu^{2}$, and we do not need to require the
small mass approximation. Then, the free energy can be rewritten
as
 \begin{equation}
 \label{TfreeEf0L}
 F = - \frac{2}{3\pi} \int^{r_H+\epsilon}_{r_H} dr
 \frac{r^2}{f^2} \int^{\infty}_{0} d\omega
 \frac{{\omega}^3}{(e^{\beta \omega} -1)\left(1+\frac{\alpha\omega^2}{f}\right)^3} .
 \end{equation}
Then, from $F$ in Equation  (\ref{TfreeEf0L}), one can find the
entropy as
\begin{eqnarray}
\label{Aentropy0L}
 S = \frac{\beta^2_H}{6\pi}
          \int^{r_H +\epsilon}_{r_H} dr \frac{r^2}{f^2}
          \int^{\infty}_{0} d\omega
          \frac{{\omega}^4 }{\sinh^2(\frac{\beta_H}{2}\omega)\left(1+\frac{\alpha\omega^2}{f}\right)^3}.
\end{eqnarray}
Introducing $x=\beta\omega/2$, the entropy can be rewritten as
 \be
  S = \frac{16}{3\pi\beta^3_H}
          \int^{r_H +\epsilon}_{r_H} dr \frac{r^2}{f^2}
          \int^{\infty}_{0} dx
          \frac{x^4}{\sinh^2(x)\left(1+\frac{4\alpha x^2}{\beta^2_H f}\right)^3}.
 \ee
Making use of the inequality
 \be
 \sinh^2(x)\ge x^2,
 \ee
and after performing $\omega$ integration, one can obtain
 \begin{eqnarray}
 S &<& \frac{16}{3\pi\beta^3_H}
          \int^{r_H +\epsilon}_{r_H} dr \frac{r^2}{f^2}
          \int^{\infty}_{0} dx
          \frac{x^2}{\left(1+\frac{4\alpha x^2}{\beta^2_H f}\right)^3}\nonumber\\
   &=& \frac{1}{24\alpha^{3/2}}
          \int^{r_H +\epsilon}_{r_H} dr \frac{r^2}{\sqrt{f}}.
 \end{eqnarray}
Since we are only interested in the contribution from just the
vicinity near the horizon in the range $(r_H,r_H+\epsilon)$, this
integration finally becomes
 \be
 S<\frac{1}{24\pi\alpha^{3/2}}\sqrt{\frac{2\epsilon}{\kappa_H}}\left(\frac{A}{4}\right)+{\cal O}(\epsilon^{3/2}).
 \ee
In the leading-order GUP correction, the minimum length is given
by Equation  (\ref{minL}), and by identifying the invariant length
with this, we have
 \be
 S<\frac{1}{24\pi\alpha}\left(\frac{A}{4}\right).
 \ee
When we choose the GUP parameter $\alpha$ as
 \be
 \alpha=\frac{1}{24\pi},
 \ee
we can finally obtain the entropy as
 \be\label{Lentropy}
 S=\frac{A}{4}.
 \ee
As a result, we have obtained the Bekenstein--Hawking entropy
satisfying the area law exactly.

It is appropriate to comment that the GUP parameter $\alpha$ is
$\alpha\approx 0.0133$. On the other hand, in Ref.
\cite{Li:2002xb}, the GUP parameter $\lambda$, which is the same
as our $\alpha$, was $\lambda=\frac{3}{4\pi}\approx 0.2387$.
Therefore, the correction in this paper is much better than the
ones in \cite{Li:2002xb} and stricter than in \cite{Kim:2007if}
where $\lambda=\frac{\sqrt{e}}{6\sqrt{2}\pi^{3/2}}\approx 0.0349$.

\subsection{All-Order GUP Correction}

In the case of all-order GUP correction in the Planck length, the
number of quantum states associated with the radial mode is given
by
\begin{eqnarray}
\label{Tnqs} n(\omega) = \int dn = \frac{2}{3\pi}\int_{r_H} dr
\frac{r^2}{\sqrt{f}} \left(\frac{{\omega}^2}{f}-
\mu^{2}\right)^{\frac{3}{2}} e^{-\alpha (\frac{{\omega}^2}{f}-
\mu^{2})}.
\end{eqnarray}
It is important to note that $n(\omega)$ remains finite at the
horizon without the need for any artificial cutoff, due to the
presence of the exponential suppression term
$e^{-\alpha\omega^2/f}$ induced by the GUP.

The free energy of a thermal ensemble of scalar fields is then
\begin{eqnarray}
\label{TfreeE}
 F = - \frac{2}{3\pi} \int_{r_H} dr \frac{r^2}{\sqrt{f}}
   \int^{\infty}_{\mu\sqrt{f}} d\omega~
    \frac{\left(\frac{{\omega}^2}{f}- \mu^{2}\right)^{\frac{3}{2}}}{e^{\beta \omega} -1}e^{- \alpha(\frac{{\omega}^2}{f}- \mu^{2})}.
\end{eqnarray}
Near the event horizon, the dominant contribution to the free
energy simplifies to
 \begin{equation}
 \label{TfreeEf0A}
 F = - \frac{2}{3\pi} \int^{r_H+\epsilon}_{r_H} dr
 \frac{r^2}{f^2} \int^{\infty}_{0} d\omega
 \frac{{\omega}^3}{e^{\beta \omega} -1} e^{- {\frac{\alpha{\omega}^2}{f}}}.
 \end{equation}
After integrating over $\omega$, the entropy can be expressed as
\begin{eqnarray}
\label{Aentropy0A}
 S = \frac{\beta^2_H}{6\pi}
          \int^{r_H +\epsilon}_{r_H} dr \frac{r^2}{f^2}
          \int^{\infty}_{0} d\omega\frac{{\omega}^4 }{\sinh^2(\frac{\beta_H}{2}\omega)}
           e^{-  {\frac{\alpha{\omega}^2}{f}}}.
\end{eqnarray}
By introducing the substitution $x=\sqrt{\alpha}\omega$, this
becomes
\begin{eqnarray}
\label{Aentropyx}
  S = \frac{\beta^2_H}{6\pi\alpha^2\sqrt{\alpha}} \int^{\infty}_{0}
          dx \frac{{x}^4}{\sinh^2(\frac{\beta_H}{2\sqrt{\alpha}}x)}\Lambda(x,\epsilon),
\end{eqnarray}
where
\begin{eqnarray}
\label{RfreeEf}
 \Lambda(x,\epsilon)  \equiv  \int^{r_H + \epsilon}_{r_H} dr~\frac{r^2}{f^2}~e^{-\frac{x^2}{f}}.
\end{eqnarray}
Focusing on the near-horizon region, Equation  (\ref{RfreeEf})
reduces to
\begin{equation}
\label{RfreeEf1}
 \Lambda(x,\epsilon) \approx  \int^{r_H + \epsilon}_{r_H} dr~
 \frac{r^2}{[2\kappa_H(r-r_H)]^2}~e^{-\frac{x^2}{2\kappa_H(r-r_H)}}.
\end{equation}
This integral can be evaluated exactly by substituting
$t=x^2/2\kappa_H(r-r_H)$, yielding
\begin{eqnarray}
 \Lambda(x,\epsilon)&=&\frac{1}{2\kappa_H x^2}\int^{\infty}_{\frac{x^2}{2\kappa_H\epsilon}} dt
          \left(r^2_H +\frac{r_Hx^2}{\kappa_H t}+\frac{x^4}{4\kappa^2_H t^2}\right)e^{-t}\nonumber\\
                    &=& \frac{r^2_H}{2\kappa_H x^2}\Gamma(1,\frac{x^2}{2\kappa_H\epsilon})
                        +\frac{r_H}{2\kappa^2_H}\Gamma(0,\frac{x^2}{2\kappa_H\epsilon})
                        +\frac{x^2}{8\kappa^3_H}\Gamma(-1,\frac{x^2}{2\kappa_H\epsilon}),
\end{eqnarray}
where the incomplete Gamma function is defined as
\begin{equation}
\Gamma(a,z)=\int^\infty_z dt~ t^{a-1}e^{-t}.
\end{equation}
The all-order GUP-corrected entropy can thus be written as
\begin{eqnarray}
\label{Aentropy1}
 S = S_1 + S_2 + S_3
\end{eqnarray}
where
\begin{eqnarray}
   S_1 &=& \frac{\beta^2_H r^2_H}{12\pi \alpha^2\sqrt{\alpha}\kappa_H}
      \int^{\infty}_{0} dx \frac{{x}^2}{\sinh^2(\frac{\beta_Hx}{2\sqrt{\alpha}})}\Gamma(1,\frac{x^2}{2\kappa_H\epsilon}),\\
   S_2 &=& \frac{\beta^2_H r_H}{12\pi \alpha^2\sqrt{\alpha}\kappa^2_H}
      \int^{\infty}_{0} dx \frac{{x}^4}{\sinh^2(\frac{\beta_H x}{2\sqrt{\alpha}})}\Gamma(0,\frac{x^2}{2\kappa_H\epsilon}),\\
   S_3 &=&  \frac{\beta^2_H}{48\pi \alpha^2\sqrt{\alpha}\kappa^3_H}
     \int^{\infty}_{0} dx
      \frac{{x}^6}{\sinh^2(\frac{\beta_Hx}{2\sqrt{\alpha}})}\Gamma(-1,\frac{x^2}{2\kappa_H\epsilon}).
\end{eqnarray}
Redefining $y=\frac{\beta_H x}{2\sqrt{\alpha}}$ and using the
minimum length (\ref{minLAO}) with $\beta_H \kappa_H=2\pi$, these
terms become
\begin{eqnarray}
\label{Fentropy1}
 S_1 &=& \frac{r^2_H}{3\pi^2 \alpha}
      \int^{\infty}_{0} dy \frac{{y}^2}{\sinh^2y}\Gamma(1,\frac{2y^2}{\pi^2e}),\\
 S_2 &=& \frac{r_H\kappa_H}{3\pi^4}
      \int^{\infty}_{0} dy \frac{{y}^4}{\sinh^2y}\Gamma(0,\frac{2y^2}{\pi^2 e}),\\
 S_3 &=& \frac{\alpha\kappa^2_H}{12\pi^6}
      \int^{\infty}_{0} dy \frac{{y}^6}{\sinh^2y}\Gamma(-1,\frac{2y^2}{\pi^2 e}).
\end{eqnarray}
Evaluating these integrals yields
\begin{eqnarray}
\label{delta1}
 \delta_1 &\equiv& \int^{\infty}_{0} dy \frac{{y}^2\Gamma(1,\frac{2y^2}{\pi^2 e})}{\sinh^2y}\approx 1.4509,\\
 \delta_2 &\equiv& \int^{\infty}_{0} dy \frac{{y}^4\Gamma(0,\frac{2y^2}{\pi^2 e})}{\sinh^2y}\approx 3.0709,\\
 \delta_3 &\equiv& \int^{\infty}_{0} dy \frac{{y}^6\Gamma(-1,\frac{2y^2}{\pi^2 e})}{\sinh^2y}\approx 18.4609.
\end{eqnarray}
Thus, the entropy can be expressed as follows
\begin{eqnarray}
\label{F1entropy1}
 S = \frac{\delta_1}{3\pi^3\alpha}\left(\frac{A}{4}\right)
      +\frac{\delta_2}{3\pi^4}r_H\kappa_H
      +\frac{\delta_3}{12\pi^6}\alpha\kappa^2_H.
\end{eqnarray}
Now, by using the surface gravity $\kappa_C$ from Equation
(\ref{kappaC}) for the all-order GUP correction and considering
the small $\alpha$ limit, we can simplify this to
 \be
 S\simeq\frac{4\delta_1M^2}{3\pi^2\alpha}
         +\frac{\delta_2}{6\pi^4}
         +\frac{(2\pi^2\delta_2+\delta_3)\alpha}{192\pi^6M^2}
         +\frac{(5\pi^2\delta_2+2\delta_3)\alpha^2}{3072\pi^6M^4}+{\cal O}(\alpha^3).
 \ee
In terms of the surface area $A=16\pi M^2$, this expression can be
further rewritten as
 \be
 S\simeq\frac{\delta_1}{3\pi^3\alpha}\left(\frac{A}{4}\right)
    +\frac{\alpha}{6\pi^3}\left(\delta_2+\frac{\delta_3}{2\pi^2}\right)A^{-1}
    +\frac{5\alpha^2}{12\pi^2}\left(\delta_2+\frac{2\delta_3}{5\pi^2}\right)A^{-2}
    +{\cal O}\left(\alpha^3A^{-3}\right).
 \ee
This formulation includes a constant term and reveals that, as the
order of $\alpha$ increases, the terms grow larger inversely with
respect to the surface area $A$. This property, except for a
logarithmic term, exhibits a characteristic of
quantum-gravity-corrected black hole entropy
\cite{Medved:2005vw,Arzano:2005rs}.

When we choose the GUP parameter $\alpha$ as
\begin{equation}
 \alpha=\frac{\delta_1}{3\pi^3}\approx 0.0156,
\end{equation}
we arrive at the final entropy expression:
\begin{eqnarray}
\label{F2entropy1}
 S \simeq \frac{A}{4}+c_1 A^{-1}+c_2 A^{-2} + {\cal O}(A^{-3}),
\end{eqnarray}
where $c_1=3.3589\times 10^{-4}$ and $c_2=3.9223\times 10^{-5}$.
Thus, through the all-order GUP correction in the effective
metric, we have finally obtained an entropy expression that
upholds the area law, along with correction terms that are
inversely proportional to the surface area.

\section{Discussion}\label{s6}

In this paper, we have studied effective metrics derived from
leading-order GUP-corrected temperature and entropy and extended
it to effective metric concerning  all-order GUP correction in
 Planck length. {Among these effective metrics,
the approach based on the leading-order GUP-corrected temperature
represents a specific limit of the all-order GUP-corrected
temperature case, while the one based on the leading-order
GUP-corrected entropy serves as an alternative construction within
the GUP framework. Each approach is applicable in different
regimes of quantum gravity corrections and is dependent on the
initial assumptions made for the metric construction.} Despite
introducing three distinct GUP-corrected effective metrics, the
statistical entropy calculated via t' Hooft's brick wall method
consistently recovers the Bekenstein--Hawking area law, which
shows the universality of the area law even in quantum-corrected
spacetimes.

We have also shown that the divergent near-horizon contribution to
entropy is regulated by a UV cutoff $h$, which differs for each
effective metric, as shown in Equations (\ref{ha}), (\ref{hb}) and
(\ref{hc}). These cutoffs ensure the entropy matches the area law,
with all reducing to the Schwarzschild case as $\alpha\rightarrow
0$. On the other hand, to eliminate coordinate dependence, the UV
cutoff $h$ is replaced with the invariant distance $l^2_{\rm
inv}$, and remarkably, all three metrics yield the same invariant
distance
 \be
 l_{\rm inv}=\frac{1}{\sqrt{90\pi}},
 \ee
confirming the regularization's physical consistency across
different effective metrics and the robustness of the area law
under GUP modifications.

These results are further confirmed by explicitly counting quantum
states in the vicinity of the event horizon without introducing
any artificial cutoff. Both the leading-order and all-order GUP
corrections naturally regularize the density of states, yielding
finite entropy and reproducing the area law exactly when the GUP
parameter is appropriately chosen. The analysis, particularly in
all-order GUP-corrected temperature, shows that subleading
corrections, including constant and inverse-area terms, arise in
the entropy expansion, reflecting the quantum gravitational
structure encoded by the GUP. These findings collectively
demonstrate that the use of full, non-perturbative GUP corrections
in effective metrics provides a consistent and physically
meaningful framework for understanding black hole thermodynamics
and that the area law remains universal even in the presence of
quantum gravity effects.

For further study, it would be interesting to extend the analysis
to {the extended uncertainty principle and the generalized
extended uncertainty principle. These alternative frameworks
introduce large-distance corrections and new phenomenological
features, particularly in cosmological or (A)dS backgrounds. They
represent an important direction for a broader understanding of
black hole thermodynamics. Additionally, it would be valuable to
investigate} rotating or charged black holes, as well as those in
higher dimensions. Such studies will reinforce the connection
between quantum gravity phenomenology and black hole
thermodynamics, emphasizing the resilience of the area law under
GUP modifications.

\vspace{+6pt}

\acknowledgments{S.-T.H. was supported by Basic Science Research
Program through the National Research Foundation of Korea funded
by the Ministry of Education, NRF-2019R1I1A1A01058449. Y.-W.K. was
supported by the National Research Foundation of Korea (NRF) grant
funded by the Korea government (MSIT) (No. 2020R1H1A2102242).}


\end{document}